\documentclass[
twocolumn,
]{ceurart}
\usepackage{amsmath,amssymb,amsfonts}
\renewcommand{\sf}[1]{\textsf{\textup{#1}}}              
\begin{document}

\title{Bridging OLAP and RAG: A Multidimensional Approach to the Design of Corpus Partitioning}

\author[1]{Dario Maio}[%
orcid=ID 
0000-0002-0094-0022,
email=dario.maio@unibo.it,
]
\cormark[1]

\author[1]{Stefano Rizzi}[%
orcid=0000-0002-4617-217X,
email=stefano.rizzi@unibo.it,
]
\address[1]{DISI - University of Bologna, Viale Risorgimento, 2, Bologna, 40136 Italy}
\cortext[1]{Corresponding author.}

\begin{abstract}
Retrieval-Augmented Generation (RAG) systems are increasingly deployed on large-scale document collections, often comprising millions of documents and tens of millions of text chunks. In industrial-scale retrieval platforms, scalability is typically addressed through horizontal sharding and a combination of Approximate Nearest-Neighbor search, hybrid indexing, and optimized metadata filtering. Although effective from an efficiency perspective, these mechanisms rely on bottom-up, similarity-driven organization and lack a conceptual rationale for corpus partitioning. In this paper, we claim that the design of large-scale RAG systems may benefit from the combination of two orthogonal strategies: semantic clustering, which optimizes locality in embedding space, and multidimensional partitioning, which governs where retrieval should occur based on conceptual dimensions such as time and organizational context. Although such dimensions are already implicitly present in current systems, they are used in an ad hoc and poorly structured manner. We propose the Dimensional Fact Model (DFM) as a conceptual framework to guide the design of multidimensional partitions for RAG corpora. The DFM provides a principled way to reason about facts, dimensions, hierarchies, and granularity in retrieval-oriented settings. This framework naturally supports hierarchical routing and controlled fallback strategies, ensuring that retrieval remains robust even in the presence of incomplete metadata, while transforming the search process from a 'black-box' similarity matching into a governable and deterministic workflow. This work is intended as a position paper; its goal is to bridge the gap between OLAP-style multidimensional modeling and modern RAG architectures, and to stimulate further research on principled, explainable, and governable retrieval strategies at scale.
\end{abstract}


\maketitle

\section{Introduction}\label{sec:intro}

In recent years, advances in transformer-based architectures have enabled Large Language Models (LLMs) to achieve remarkable capabilities in natural language understanding and generation \cite{vaswani2017attention,brown2020language}, supporting a wide range of applications that require flexible access to large and heterogeneous knowledge sources. These models are increasingly regarded as foundation models, providing general-purpose linguistic capabilities while also raising challenges related to knowledge freshness, control, and integration with external data sources \cite{bommasani2021opportunities}.
However, LLMs are inherently parametric and static, and their knowledge is limited to what has been encoded during training. To address this limitation, Retrieval-Augmented Generation (RAG) has emerged as a dominant architectural paradigm, combining generative models with external retrieval mechanisms over large document collections \cite{lewis2020rag}.

In a RAG system, user queries are used to retrieve relevant textual fragments from a corpus, which are then provided as contextual input to the LLM during generation. This approach allows systems to ground responses in up-to-date and domain-specific information, and has become the de facto standard for deploying LLM-based applications over enterprise-scale data. As RAG systems are increasingly applied to collections comprising millions of documents and tens or hundreds of millions of text chunks, issues of scalability, robustness, and governability of retrieval become central.

Current large-scale RAG architectures primarily rely on similarity-based retrieval over vector representations, often combined with hybrid indexing and approximate nearest-neighbor search techniques \cite{gao2024retrieval,chen2024benchmarking}. Modern industrial RAG systems already combine vector-based similarity search with clustering and metadata filtering strategies; however, these mechanisms are primarily driven by bottom-up semantic similarity and do not provide an explicit conceptual rationale for corpus partitioning.

Scalability in industrial RAG systems is typically achieved through horizontal sharding, hybrid indexing strategies, and metadata-based filtering used to restrict the retrieval space when possible. While effective from an efficiency standpoint, these solutions are largely implementation-driven: corpus organization is dominated by embedding-space locality, and logical constraints on where retrieval should occur are handled in an ad hoc and weakly structured manner.

In parallel, the problem of organizing and querying large volumes of data has been long addressed in Online Analytical Processing (OLAP) systems and data warehouses \cite{kimball2013dw,golfarelli2009dwbook}. Multidimensional modeling provides a principled framework for representing facts, dimensions, hierarchies, and levels of granularity, supporting systematic reasoning about data organization. In particular, the Dimensional Fact Model (DFM) operates at a conceptual level, guiding design decisions while remaining agnostic with respect to implementation details such as physical storage and execution strategies \cite{golfarelli1998dfm}. 

In this paper, we argue that large-scale RAG systems would benefit from a similar conceptual design layer. We propose to reinterpret the DFM as a framework for the multidimensional partitioning of unstructured document corpora used in RAG. In our view, effective RAG architectures should explicitly separate two orthogonal concerns: (i) the \emph{multidimensional partitioning} of the corpus along domain-relevant dimensions, used to guide routing decisions and determine where retrieval should take place; and (ii) semantic clustering and sub-clustering applied within each multidimensional partition, used to support similarity-based retrieval over vector representations. Note that clustering and sub-clustering are cited solely as representative examples of semantic organization strategies that may be adopted within multidimensional partitions. The proposed framework does not impose any specific retrieval logic, nor does it interfere with the internal filtering, indexing, or execution strategies adopted by existing industrial retrieval systems. Rather, it provides a principled way to govern retrieval by making conceptual assumptions explicit and controllable.

The proposed approach is intentionally agnostic with respect to underlying system architectures. Multidimensional partitions are defined independently of physical sharding, although in practical deployments a mapping between multidimensional partitions and physical shards naturally exists and follows standard scalability and load-balancing practices. Similarly, the presence or absence of a data warehouse is orthogonal to the applicability of the framework. When a data warehouse or other structured repositories are available, they facilitate the design of dimensions as their multidimensional schemata may be available. When such infrastructures are absent, the DFM remains a purely conceptual tool for reasoning about corpus organization and retrieval strategies. In both cases, the problem of metadata extraction and management is inherent to large-scale retrieval systems and is not introduced by the proposed framework.

This work is a position paper and addresses methodological issues. Rather than proposing new retrieval algorithms or experimental evaluations, it introduces a conceptual perspective that complements existing RAG architectures, with the goal of stimulating discussion on principled, explainable, and governable design strategies for retrieval over massive unstructured corpora.
The contributions of this work can be summarized as follows:

\begin{itemize}

\item We identify a conceptual gap in current RAG architectures, where corpus partitioning and metadata-based routing are critical for scalability and robustness, yet are addressed in an ad hoc, implementation-driven manner without an explicit conceptual design layer.

\item We argue that effective large-scale RAG systems require the explicit combination of two orthogonal organizational principles: semantic clustering for embedding-space locality, and multidimensional partitioning for governing where retrieval should occur based on domain-relevant dimensions.

\item We propose the DFM as a conceptual framework to guide multidimensional partitioning of large unstructured corpora for retrieval purposes. 

\item We show that a DFM-based design naturally enables hierarchical routing and controlled fallback strategies during retrieval, supporting deterministic, explainable, and governable retrieval workflows even in the presence of incomplete or partially missing metadata.

\item We present an illustrative example from a complex legal domain, involving collections of judicial decisions or contractual documents at industrial scale, to demonstrate how multidimensional modeling can structure the retrieval space along meaningful dimensions without constraining underlying similarity-based retrieval mechanisms.

\end{itemize}


The remainder of the paper is organized as follows. 
Section~\ref{sec:related} reviews related work on RAG architectures, metadata-driven retrieval, and multidimensional modeling.  Section~\ref{sec:dfm-rag} introduces the DFM and discusses its reinterpretation in a retrieval-oriented setting. 
Section~\ref{sec:example} presents illustrative examples drawn from complex application domains to motivate the proposed approach. Finally, Section~\ref{sec:discussion} discusses the implications of the framework, its limitations, and some open directions for future research.

\section{Related Work}\label{sec:related}

\subsection{Conceptual Modeling and LLMs in OLAP}

Conceptual modeling has long played a central role in the design of data warehouses and OLAP systems. The multidimensional model provides a principled framework for representing facts, dimensions, hierarchies, and granularity at a conceptual level, supporting systematic reasoning about aggregation paths, drill-down and roll-up operations, and alignment between analytical requirements and physical implementations \cite{kimball2013dw,golfarelli2009dwbook}. The DFM formalizes these concepts at the conceptual level and has been widely adopted as a reference model for data warehouse design \cite{golfarelli1998dfm}.
Recent studies have investigated the role of LLMs in the conceptual design of OLAP systems. Rizzi \cite{rizzi2025chatgpt} explores the use of ChatGPT to refine draft conceptual schemata in supply-driven design of multidimensional cubes, while Rizzi et al.\ \cite{rizzi2025llmmd} provide an empirical investigation of the potential and limitations of LLMs in assisting multidimensional modeling tasks. These works demonstrate that LLMs can effectively operate at the conceptual level of OLAP design.

Research on natural language interfaces to databases has a long tradition in the data management community, with the goal of translating user questions into formal query languages such as SQL. Recent advances in neural and transformer-based language models have significantly revitalized this area, leading to modern Text-to-SQL and schema-aware semantic parsing approaches that achieve strong performance on complex and cross-domain benchmarks such as \textsc{Spider} \cite{yu2018spider}. This line of work primarily aims at improving accessibility to structured data by mapping user intents to queries over an existing database schema.
More recently, LLMs have been explored as flexible interfaces for analytical systems, supporting conversational interaction and explanation over OLAP and Business Intelligence platforms \cite{DBLP:conf/er/BimonteR25}. In these approaches, LLMs act as front-ends for querying and exploring multidimensional data, while assuming that the underlying OLAP schema and data organization are given.

Despite its maturity and widespread use in analytical systems, the multidimensional model has not been explicitly applied to the design of retrieval structures for large-scale unstructured document collections. In current RAG architectures, conceptual modeling is largely absent: retrieval units, metadata, and partitions are defined in an ad hoc manner, and semantic clustering is relied upon as the primary organizational principle.

\subsection{Partitioning and Metadata in RAG}

Several works in the RAG literature have recognized the importance of structuring the retrieval space to improve generation quality. Wang et al.\ \cite{wang-etal-2024-rag} propose a multi-partition RAG architecture (M-RAG), where multiple partitions act as basic units for retrieval and generation, showing that partitioning can significantly improve performance across different language generation tasks. In this approach, however, partitions are treated as operational units optimized during execution, without an explicit conceptual schema guiding their design.

Other studies explore the use of metadata to filter or restrict the retrieval space in RAG pipelines. Di Oliveira et al.\ \cite{dioliveira2025twostep} propose a two-step RAG approach in which metadata filtering precedes vector-based retrieval, demonstrating that structured metadata can effectively reduce noise and improve statistical evaluation of LLM outputs.

Finally, early attempts to integrate OLAP-style workflows with RAG architectures have been proposed. Ouafiq and Saadane \cite{ouafiq2025raolap} introduce a Retrieval-Augmented OLAP architecture in which multidimensional analytical structures are used to support the reasoning process of generative models. While these approaches confirm the relevance of multidimensional organization and metadata-driven partitioning in retrieval systems, they do not provide a general conceptual framework for the design of partitions over unstructured document corpora.

\subsection{RAG with Structured Data Sources}
While the previous works focus on partitioning and metadata-driven control of the retrieval space, another line of research investigates the integration of structured data sources into RAG pipelines. Several works combine RAG with relational databases, knowledge bases, or graph databases, enabling LLMs to retrieve both unstructured text and structured facts during answer generation \cite{lewis2020rag,yao2023react,schick2023toolformer}. In these systems, structured queries are typically executed via external tools or dedicated query engines, while unstructured retrieval relies on vector similarity search over text passages.

Hybrid architectures that interleave semantic retrieval with database querying have been proposed to improve factual accuracy, reduce hallucinations, and support multi-step reasoning \cite{yao2023react,schick2023toolformer}. However, in most of these approaches the unstructured corpus is treated as a flat or weakly structured collection, possibly augmented with simple metadata filters. The design of partitions for the text corpus is generally left to heuristic decisions or similarity-based clustering, without an explicit conceptual model guiding partitioning choices.

\subsection{Positioning of This Work}

In contrast to the aforementioned lines of research, this paper focuses on the conceptual design of multidimensional partitions for retrieval over unstructured corpora. Rather than introducing new retrieval algorithms or hybrid query mechanisms, we argue that principles from multidimensional modeling can provide a formal design layer for large-scale RAG systems that is currently missing. To the best of our knowledge, this is the first work that explicitly proposes the DFM as a conceptual framework to reason about multidimensional partitioning, hierarchical routing, and fallback strategies in RAG.

\section{Applying the DFM to Retrieval-Oriented Partitioning}\label{sec:dfm-rag}

In OLAP systems, the DFM provides a conceptual representation of analytical data in terms of facts, dimensions, hierarchies, and levels of granularity, independently of physical storage and query execution strategies \cite{golfarelli2009dwbook}. 
In this context, a fact represents a business event of interest, such as an invoice or a sales transaction, described through multiple dimensions such as time, customer, product, and location. 

Building on this well-established abstraction, we reinterpret the basic retrieval unit in a RAG system (e.g., a text chunk) as the analogue of a fact, while dimensions correspond to domain-relevant attributes that characterize documents and guide retrieval decisions, such as time, document type, jurisdiction, or organizational context. Hierarchies within dimensions capture meaningful levels of abstraction and support progressive refinement of the retrieval space. Importantly, the proposed model does not prescribe how such dimensions are physically indexed or stored, but only how they are conceptually organized and combined. 
In other words, in this work, we reinterpret these concepts in the context of RAG, where the primary goal is not aggregation but the governable selection of relevant portions of a large unstructured corpus.

Within this framework, multidimensional partitioning and semantic retrieval play complementary roles. Multidimensional partitions, defined along one or more dimensions, are used to route queries and determine where retrieval should occur. Semantic clustering and sub-clustering are then applied within each multidimensional partition to support similarity-based retrieval over vector representations. This separation allows embedding-based methods to operate locally, while global retrieval behavior remains governed by explicit conceptual choices.

The model explicitly accommodates incomplete or missing metadata by allowing dedicated fallback partitions that collect documents whose dimensional values cannot be reliably determined. From a multidimensional perspective, such partitions correspond to facts with optional dimensional coordinates and are treated as first-class elements of conceptual design, ensuring robustness without relying on implicit or ad hoc behaviors.

Overall, the proposed reinterpretation positions the DFM as a conceptual design layer for large-scale RAG systems, enabling principled reasoning about corpus organization, routing strategies, and retrieval scope, while remaining fully agnostic with respect to underlying architectures and execution mechanisms. By making partitioning assumptions explicit, the DFM provides a stable conceptual reference for governing the evolution of routing, clustering strategies, and fallback policies as the corpus and usage patterns change over time.

\section{Illustrative Example}\label{sec:example}

To illustrate the proposed approach, we consider a large-scale legal domain scenario involving collections of judicial decisions or contractual documents. Such corpora are characterized by high structural complexity, strong domain semantics, and industrial-scale volumes, often comprising millions of documents and hundreds of millions of text chunks when processed for RAG.
In the absence of a multidimensional perspective, retrieval in such settings is typically performed over a flat or weakly-structured corpus, possibly augmented with ad hoc metadata filters. As a result, semantically similar fragments originating from different jurisdictions, procedural levels, or temporal contexts may be retrieved together, despite being conceptually incompatible for a given information need.

Using a multidimensional perspective, the corpus can instead be logically partitioned along legally-meaningful dimensions, such as time, jurisdiction, procedural posture, and document type. These dimensions define partitions that guide query routing and determine where retrieval should occur. For instance, a query concerning recent appellate decisions in a specific jurisdiction can be routed to the corresponding multidimensional partitions, while documents lacking reliable temporal or jurisdictional metadata are explicitly assigned to a dedicated fallback partition.
Within each multidimensional partition, semantic clustering and sub-clustering strategies can be applied to organize vector representations and support similarity-based retrieval at different granularities. In this way, embedding-based methods operate within conceptually coherent subsets of the corpus, while global control over the retrieval space is preserved by the multidimensional partitioning layer.

\begin{figure}
    \centering
    \includegraphics[width=\columnwidth]{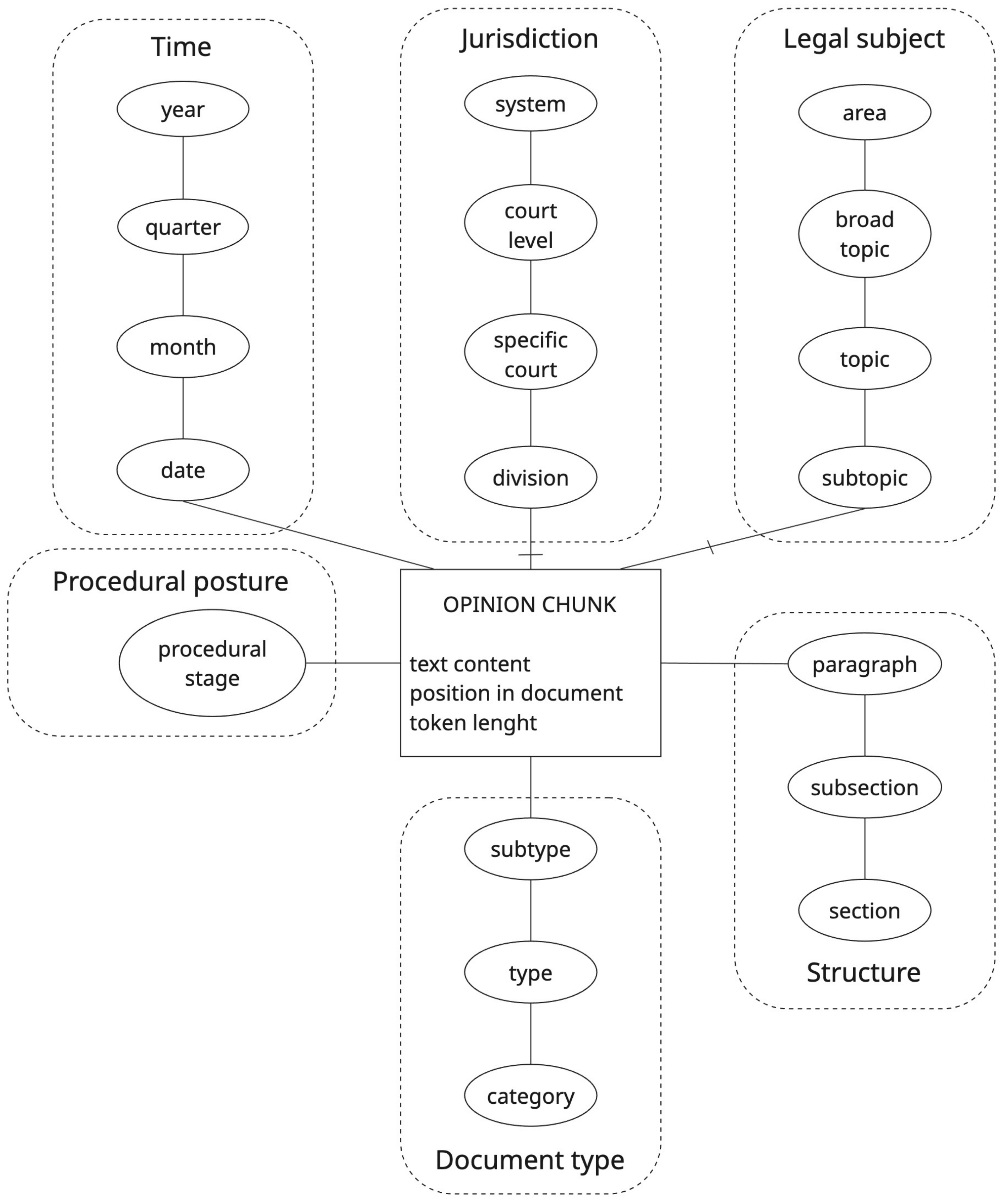}
    \caption{DFM schema for judicial opinion chunks in the legal domain. The box represents the basic retrieval unit with its quantitative measures; dashed areas represent dimensions (e.g., \sf{Structure}), while circles represent their hierarchical levels from the finest ones (those attached to the box, e.g., \sf{paragraph}) to the coarsest ones (the leaves, e.g., \sf{section}). Dashes on arcs model optional dimensions.}
    \label{fig:legal-dfm}
\end{figure}

Figure~\ref{fig:legal-dfm} illustrates a possible DFM schema for judicial opinion chunks in the context of the United States legal system.
The basic retrieval unit corresponds to an \emph{opinion chunk}, i.e., a portion of a judicial decision obtained by segmenting court opinions into manageable text fragments for RAG. Opinion chunks are derived from documents such as judicial opinions, orders, or memoranda, typically available in formats such as PDF or HTML and processed into text chunks during ingestion. They correspond to atomic units of legal discourse that may be retrieved and provided as contextual evidence to a language model.
Each opinion chunk is characterized by multiple orthogonal dimensions commonly used in legal research and analysis, including:

\begin{itemize}
\item \sf{Jurisdiction}, capturing the institutional context of the decision in terms of (from the finest to the coarsest level) \sf{division}, \sf{specific court}, \sf{court level} (e.g., 'supreme' and 'appeals district'), and \sf{system}  (either 'federal' or 'state');

\item \sf{Legal subject}, which gives a taxonomic description of the substantive legal area addressed by the opinion with levels such as \sf{area} ('civil', 'criminal', 'constitutional') and \sf{broad topic} (e.g. 'contracts' and 'torts');

\item \sf{Document type}, distinguishing different kinds of legal documents and opinions mainly in terms of \sf{category} (e.g., 'opinion' and 'order') and \sf{type} (e.g., 'majority' and 'dissent');

\item \sf{Procedural posture}, describing the procedural context in which the decision was issued (e.g., 'trial' and 'appeal');

\item \sf{Structure}, identifying the rhetorical or functional role of the chunk within the opinion (e.g., 'facts', 'legal analysis', 'holding' for level \sf{section}).
\end{itemize}
Opinion chunks whose dimensional values cannot be reliably determined are explicitly associated with dedicated fallback partitions; in the DFM, these are represented as optional dimensions (in our example, \sf{Legal subject} and \sf{Jurisdiction}).

Note that the proposed schema is intended solely as a conceptual reference and does not prescribe any specific physical storage, indexing strategy, or system architecture. The mapping between multidimensional partitions and physical design choices---such as shards, indexes, and cluster organizations---is deployment-dependent and remains outside the scope of this work.

Within this framework, multidimensional information is used to guide query routing and determine the scope of retrieval before any similarity-based search is performed. Queries are analyzed to extract dimensional constraints —--explicit or implicit--— which are then used to select the relevant multidimensional partitions of the corpus.
For example, a query seeking recent Supreme Court decisions on a specific legal doctrine may activate constraints on the \sf{Time}, \sf{Jurisdiction}, and \sf{Legal subject} dimensions, routing retrieval to partitions corresponding to recent Supreme Court opinions addressing that doctrine. Similarly, queries aimed at comparing appellate decisions across multiple circuits may intentionally span several jurisdictional partitions, enabling structured comparison rather than conflating semantically similar but institutionally distinct decisions.

Dimensions and their hierarchical levels further allow retrieval to be focused on specific parts of opinions, such as legal analysis sections, avoiding the inclusion of factual background or procedural history when these are not relevant to the information need. When dimensional constraints are weak, ambiguous, or incomplete, fallback partitions (represented via optional dimensions) ensure that retrieval remains robust and explicitly governed rather than relying on implicit behavior.

Within each selected multidimensional partition, semantic retrieval techniques—such as vector similarity search supported by clustering or sub-clustering—can then be applied to identify the most relevant opinion chunks. In this way, dimensional routing governs \emph{where} retrieval should occur, while semantic similarity determines \emph{what} content is most relevant within that scope.



It is important to note that the proposed multidimensional partitioning does not impose restrictive access patterns on retrieval. In scenarios where users explicitly seek broad coverage, such as surveying legal precedents across multiple years, courts, or jurisdictions, queries may legitimately span multiple or all multidimensional partitions. In such cases, the proposed framework does not reduce the inherent computational cost of retrieval and is not intended to do so. Rather, it provides an explicit and controlled way to express and govern these retrieval scopes, making broad searches a deliberate design choice rather than an implicit side effect of a flat corpus organization.

A similar rationale applies to large-scale biomedical and healthcare-related document collections, such as scientific articles, clinical guidelines, and technical reports, which are typically available in heterogeneous formats and processed into text chunks for RAG. These corpora may comprise millions of documents and hundreds of millions of chunks and are characterized by strong domain semantics and rich, yet often incomplete, metadata. Meaningful dimensions may include publication time, medical condition, type of study or document, target population, and source. Partitioning along such dimensions can guide retrieval toward conceptually appropriate subsets of the corpus (e.g., recent guidelines versus primary studies, or adult versus pediatric populations), while semantic retrieval techniques operate within each partition. As in the legal scenario, documents whose metadata cannot be reliably determined can be consistently assigned to dedicated fallback areas of the retrieval space, ensuring robustness without implicit or uncontrolled retrieval behavior.

Similarly, while certain dimensions such as time may appear universal and already widely used in practice, the contribution of the DFM lies not in the introduction of individual dimensions, but in their systematic organization, combination, and governance. By making dimensions, hierarchies, and fallback strategies explicit at the conceptual level, the proposed approach supports flexible routing strategies that adapt to different query intents, without constraining physical execution or performance optimization mechanisms.

\section{Discussion and Implications}\label{sec:discussion}

This paper does not introduce a new retrieval technique or system architecture, but rather it advocates the introduction of a conceptual perspective on the design of large-scale RAG systems, inspired by well-established principles from multidimensional modeling.
As such, its primary implications concern governability, explainability, and design clarity, rather than performance optimization. By making partitioning decisions explicit and grounding them in a multidimensional schema, the proposed framework enables systematic reasoning about where retrieval should occur, how retrieval scopes are defined, and how fallback strategies are handled.

From a query perspective, RAG systems are expected to support heterogeneous information needs, ranging from document-centric queries (e.g., lookup and document-level summarization), to retrieval-oriented queries (e.g., thematic or multi-document retrieval), per-document views, corpus-level aggregation, and multi-step reasoning queries. These different query types naturally interact with the retrieval space at different granularities and scopes. Accordingly, different classes of queries may interact with multidimensional partitioning in different ways: while some queries require minimal or no partition-based restriction, others benefit from explicit control over the retrieval scope to ensure completeness, correctness, and reproducibility of results.

An important implication of this approach is its flexibility with respect to different query intents. While multidimensional partitioning can be used to restrict retrieval to focused subsets of the corpus, it does not prevent broad queries that legitimately span multiple or all partitions, such as exploratory searches over large temporal or organizational ranges. In these cases, the computational cost of retrieval remains inherent to the task and is not reduced by the proposed framework. Instead, the framework provides an explicit and controlled way to express such retrieval scopes, avoiding implicit or unintended behavior. While performance improvements are not the primary goal of the proposed framework, queries whose information needs can be satisfied within a limited set of multidimensional partitions may naturally benefit from a reduced retrieval scope.

The proposed use of the DFM does not rely on the novelty of individual dimensions, some of which may be widely adopted in practice (e.g., time). Its contribution lies in the systematic organization of dimensions, hierarchies, and levels of granularity, as well as in the explicit treatment of incomplete or missing metadata. These aspects are often handled implicitly in existing systems, whereas the proposed approach elevates them to first-class design decisions.

Some limitations should be acknowledged. Our framework does not address the problem of metadata extraction, which remains a challenge in large-scale document processing regardless of the retrieval architecture. Moreover, it does not prescribe how multidimensional partitions should be mapped to physical shards, indexed, or executed, leaving these choices to existing industrial solutions and optimization strategies. Consequently, the framework should be viewed as complementary to, rather than a replacement for, current retrieval technologies.

Overall, we believe that the multidimensional perspective introduced in this work will be able to support more principled, transparent, and adaptable retrieval architectures, and we hope it will stimulate further research on conceptual modeling approaches for retrieval over large unstructured corpora. We encourage the development of open-source tools to support conceptual design in this direction, particularly for assisting the definition and governance of multidimensional partitions in RAG systems; when data warehouses or other structured repositories are already in place, such tools should naturally integrate with existing infrastructures rather than replace them.

\section*{Acknowledgment}
We acknowledge financial support under the PR Puglia FESR FSE+ 2021-2027 - European Regional Development Fund, Net Service project QUICK SHIELD of Puglia Region, CUP:B85H24000920007.

\bibliography{references}

\end{document}